\documentclass[sigconf]{acmart}

\usepackage{subfigure			}
\usepackage{balance}
\usepackage{soul}
\usepackage{hyperref}
\usepackage{url}
\usepackage{xcolor}
\usepackage{framed}
\usepackage{booktabs}
 \usepackage{multirow}
 \usepackage{tablefootnote}
 \usepackage{footnote}
 \usepackage{graphicx}

\begin{document}


\title{Analyzing Offline Social Engagements: An Empirical Study of Meetup Events Related to Software Development}


\fancyhead{}








\author{Abhishek Sharma}
\affiliation{%
 \institution{School of Information Systems\\ SMU, Singapore\\}
}
\email{abhisheksh.2014@smu.edu.sg}

	\author{Gede Artha Azriadi Prana}
\affiliation{%
	 \institution{School of Information Systems\\ SMU, Singapore\\}
}
\email{arthaprana.2016@smu.edu.sg}

	\author{Anamika Sawhney}
\affiliation{%
\institution{School of Information Systems\\ SMU, Singapore\\}
}

\email{anamikas@smu.edu.sg}

\author{Nachiappan Nagappan}
\affiliation{%
	\institution{Microsoft Research\\ Redmond, WA USA}
}
\email{nachin@microsoft.com}

\author{David Lo}
\affiliation{%
\institution{School of Information Systems \\SMU, Singapore}}
\email{davildo@smu.edu.sg}

\begin{abstract}

Software developers use a variety of social media channels and tools in order to keep themselves up to date, collaborate with other developers, and find projects to contribute to. Meetup is one of such social media used by software developers to organize community gatherings. Liu et al. characterized Meetup as an event-based social network (EBSN) which contains valuable \textit{offline} social interactions in addition to online interactions. 
Recently, Storey et al. found out that Meetup was one of the social channels used by developers. We in this work investigate in detail the dynamics of Meetup groups and events related to software development, which has not been done in any of the previous works.

In this work, we performed an empirical study of events and groups present on Meetup which are related to software development. First, we identified 6,317 Meetup groups related to software development and extracted 185,758 events organized by them. Then we took a statistically significant sample of 452 events (95\% confidence level with 5\% error margin) on which we performed open coding, based on which we were able to develop 9 categories of events (8 main categories +``Others''). Next, we did a popularity analysis of the categories of events and found that {\em Talks by Domain Experts, Hands-on Sessions}, and {\em Open Discussions} are the most popular categories of events organized by Meetup groups related to software development.  Our findings show that more popular categories are those where developers can learn and gain knowledge. On doing a diversity analysis of Meetup groups we found 19.82\% of the members on an average are female, which is a larger proportion as compared to numbers reported in previous studies on other social media. We also found that the groups related to technologies such as {\em blockchain } and {\em machine learning} have become more popular during last 5 years. We also built a topic association graph based on the topics. Our study results show that a lot of software community related information  is present in Meetup data, which can be used by software engineering research community for further research explorations. From a broader software development community point of view information from this new forum can be valuable to identify and understand emerging topics and associations among them which can be helpful to identify future trends as well as current best practices. 

\end{abstract}



%

\maketitle
\section{Introduction}
\label{sec.intro}

Software development has evolved into an increasingly social activity over past few decades. Social media such as social coding sites, Q\&A forums, and microblogs are used  extensively by developers for activities such as reusing other projects and tools, keeping up to date, learning new skills, connecting and collaborating with other developers, etc.~\cite{storey2014r}. Storey et al. had done a survey to understand how various social channels shape the participatory culture in software development~\cite{storey2017social}.  One of such channels touched upon in their work is Meetup\footnote{\url{https://www.meetup.com/}}. Meetup is an online social networking service, which allows people to organize events and gatherings. It allows people to form groups or communities focused on common topics of interest. The organizers of such groups can then organize off-line gatherings or events. The events that are organized range from informal congregations to formal events such as conferences.  Liu et al. characterized Meetup as an event-based social network (EBSN) which contains valuable \textit{offline} social interactions in addition to online interactions~\cite{liu2012event}. It is one of the biggest EBSNs available today with 35,324,171 members spread across 307,436 groups~\cite{aboutmeetup2018}.  




In this empirical study, taking cue from Storey et al's work 
we analyze what kinds of events are organized in Meetup groups related to software development. Storey et al. had mentioned that Meetup  was one of the channels used by developers, however they did not investigate its usage detail~\cite{storey2017social}. We wanted to investigate to uncover the categories of events organized by Meetup groups and if there are any popular topics associated with such events and groups. In this work we have empirically analyzed the usage of Meetup groups and events in detail. First, we found 6,327 Meetup groups whose associated  topics (assigned by group organizers) are related to software development. Then, from these groups we extracted a candidate data set of 185,758 events. From this candidate data set, we took a random sample of 100 events. This sample was then analyzed using the open coding methodology~\cite{saldana2015coding} in order to develop categories of events related to software development. Then based on these categories we further labeled 400 more randomly sampled events using the methodology used in ~\cite{aniche2018modern,uddin2017mining}. We decided to label more events in order to have a statistically significant sample. In the end we had 452 events labeled into some categories. The final labeled data sample is less than 500 as we dropped some events which labelers found hard to label. The sample of 452 events is statistically significant with 95\% confidence level with 5\% error margin. Based on this analysis we investigate 5 research questions. 1) What are the categories of events organized by Meetup groups related to software  development? 2) How popular is each event category? 3) How diverse are Meetup groups with respect to gender? 4) How does popularity of topics relate to Meetup groups changes over time? 5) Can we observe any topic associations among group topics?

We were able to find 9 categories of events (8 main categories +``Others'') after performing the open coding procedure. We also found that {\em Talks by Domain Experts, Hands-on Sessions,} and{ \em Open Discussions} are the most popular categories of events organized by Meetup groups related to software development. We also found that around 19.82\% of the members on Meetup groups are female, which is a higher proportion when compared to numbers reported in previous studies such as 3-9\% on Github~\cite{vasilescu2015gender,david2008community} and 7\% on Stack Overflow~\cite{vasilescu2013gender}.  Further we observed that during recent years more  Meetup groups are being formed which are associated with  topics such as {\em collaboration, ethereum, kotlin, tensorflow,} and {\em hyperledger} , whereas no or very few new groups are being formed for topics such as  {\em meteor\footnote{\url{https://www.meteor.com/}}, mapreduce, go, visual-studio,} and {\em web-performance}.  The major contributions of our work are as follows:

\vspace{-0.11cm}
\begin{enumerate}
\item We are the first to empirically investigate what kinds of events are organized by Meetup groups related to software development. We
performed an open coding procedure on 100 events and subsequent manual data labeling on 400 more events,
to group them into categories, and to find the popular categories of events.
\item We performed an analysis of how interests of software development community have changed over time by analyzing the impact of topics related to Meetup groups over a 5-year period. The impact of a topic is the percentage of groups created in a year for a topic, with respect to the total number of groups created in the same year. It is explained in detail in Section~\ref{sec:RQ3}. We also explored if any technology associations can be observed from topics.
\end{enumerate}

\vspace{-0.05cm}
\section{Background}
\label{sec.background}
In this section, we give some background of how Meetup groups  are formed and how their members organize events. The Meetup website\footnote{\url{http://www.meetup.com/}}, which was launched in 2002, works as an event scheduling and group organization tool in which members can seek, join, or create groups focusing on certain interests or activities, such as art, software development, or travel. It has been characterized as an event-based social network (EBSN)~\cite{liu2012event}, which not only contains online social interactions but also valuable offline engagement and interaction among participants. 
In order to create or join a  group on Meetup, a person first needs to register as a member on the Meetup website. The members may provide some keywords which represent their topics of interest and also their current location. This helps the website to recommend local groups related to the topics which a member has expressed interested in. Each Meetup member can be a part of one or more groups, and may hold different positions in each group (e.g., organizer, co-organizer, assistant organizer, event organizer, basic member, etc.).



 Members in addition to joining existing groups, can also create new groups based on the topics of their interest. During the creation of a group, the  creator (also known as``organizer'') of the group is prompted to specify the group's location (``hometown'') as well as one or more topics associated with the group. The group can be associated with one of the pre-defined Meetup categories, such as ``Arts'', ``Language \& Culture'', ``Tech'', etc.  In addition to such categories, organizer can also associate fine grained topics related to groups using keywords such as ``python", ``software development", ``machine learning", etc. A full list of topics can be found here.\footnote{\url{https://www.meetup.com/topics/}} 

\begin{figure}
	\centering
	\includegraphics[width=1\linewidth]{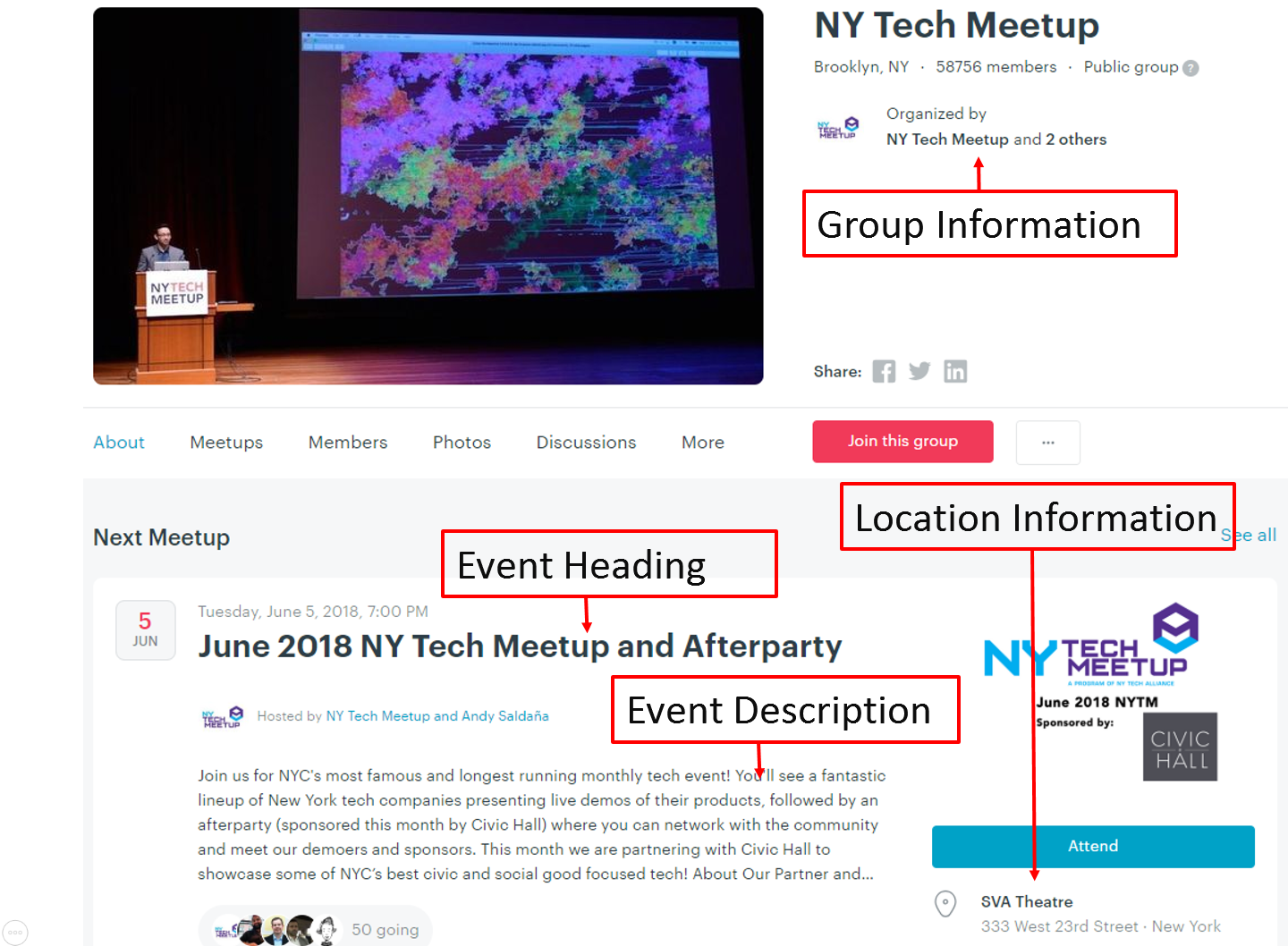}
	\vspace{-0.15cm}
	\caption{A Sample Meetup Group Related to Software Development}
	\vspace{-0.15cm}
	\label{fig:meetupgroupscreenshot}
\end{figure}

Within a Meetup group, the group organizer and the rest of the group's leadership team can plan one-off or recurring events such as a  weekly group discussions, workshops, or talks by experts. The events can require participation fees or can be complimentary. The actual content, schedule, and fee of each event is set by its organizer, and each group can have arbitrary combination of one-off and regular events of various types. The event organizers when creating an event provide a one line description as event heading, as well as a detailed description describing details such as who will be speaking at the event,  general theme of the event, etc. Group members can opt to join any event, provided the event's registration deadline has not passed and the event has not hit its attendee limit. 

Figure~\ref{fig:meetupgroupscreenshot} shows a sample Meetup group (\url{https://goo.gl/czn81K}) and some upcoming events. The top of the page shows \textit{ group information} such as  group name, group location and the number of members in the group.
 The \textit{event heading} part  gives a very brief overview of the event which is then followed by \textit{event description}. The \textit{event description} describes in detail the agenda of the event.  Also, the \textit{location }and \textit{scheduling }information is present on the right side of the page. 

\section{Related Work}
\label{sec.related}


%


\textbf{Social Media and Software Engineering:}  
As emphasized by Storey et al., social media has revolutionized the way software development is done~\cite{storey2014r}.  In their recent work, Storey et al. have also investigated how usages of various social and communication channels affect software development~\cite{storey2017social}. They surveyed developers about  how they use various tools such as social coding websites, question answering forums, mailing lists, micro-blogging services, chat etc.  and found out the challenges developers face in using these tools. They also briefly mentioned about Meetup being one of the channels used by software developers. The role of social networking in software development was also examined by several other prior works~\cite{begel2013social,storey2010impact}. There have been many works  which have analyzed individual sites or channels; we describe some of them below.

Stack Overflow is one of the most popular channels that has been a subject of many empirical studies~\cite{jones2009conversations,barua2014developers,vasilescu2013stackoverflow,linares2013exploratory,linares2014api,calefato2015mining}. Also, Stack Overflow data has been used to build various automated tools to support software development~\cite{treude2016augmenting,uddin2017automatic,tian2017apibot,murgia2016among,chen2016mining}. Another channel that has received a lot of attention  is GitHub, a social coding website.  There have been various works studying developer behaviors and practices by analyzing data in GitHub~\cite{blincoeIST15,kalliamvakouICSE15,thung2013network,VasilescuBXCDDF16,CasalnuovoDOFR15,ray2014large,rastogi2016forking,rastogi2018relationship}. There have also been studies that investigate other channels such as Twitter and YouTube on contents related to software development. Developers' usage of microblogs has been explored by a number of prior empirical studies~\cite{Storey2014,SingerFS14,Bougie11,Wang13}. A number of other studies have built tools for analyzing Twitter content related to software development ~\cite{Tian12does,Tian14,prasetyo12,sharma2015nirmal,sharma2017harnessing,sharma2018recommending}. Analysis of how developers use screencasts was explored in ~\cite{macleod2015code}. Methods to extract relevant fragments from software development videos tutorials have been discussed in  ~\cite{ponzanelli2016codetube,ponzanelli2017automatic}. An analysis of user comments on coding video tutorials present on YouTube was performed in~\cite{poche2017analyzing}. Recently developers' usage of new channels such as {\em slack}, \textit{Reddit} and \textit{HackerNews} has also been explored~\cite{lin2016developers,aniche2018modern,chatterjee2019exploratory}

\label{sec.related.socialmedia_short}
 
\vspace{.1cm}\noindent 
 \textbf{Event Based Social Networks:} 
 Shen et al. demonstrate that Meetup events contribute to the creation of social capital ~\cite{shen2015exodus}.  Liu et al. introduced the term event-based social network (EBSN)  to categorize online services such as Meetup, Eventbrite, etc. and found that  communities in EBSNs are more cohesive than those in other types of social networks ~\cite{Liu11}. Many studies have proposed machine learning solutions for recommending events in EBSNs ~\cite{macedo2015context,lu2016location,qiao122014event,liao2016whether,pham2015general}. Most recently, Pramanik et al. proposed an algorithm that can predict Meetup group success~\cite{pramanik2016predicting}.


 \label{sec.related.ebsn}
 
\vspace{.1cm}\noindent Our work is different from previous works as we are the first to characterize events in Meetup generated by groups that are interested in software development which has not been looked into previous works related to software engineering or EBSNs.

\vspace{-0.2cm}
\section{Research Setting}
\label{sec.setting}
\begin{figure}[!htb]
	\vspace{-0.5cm}
	\begin{center}
			\includegraphics[width=3.4in]{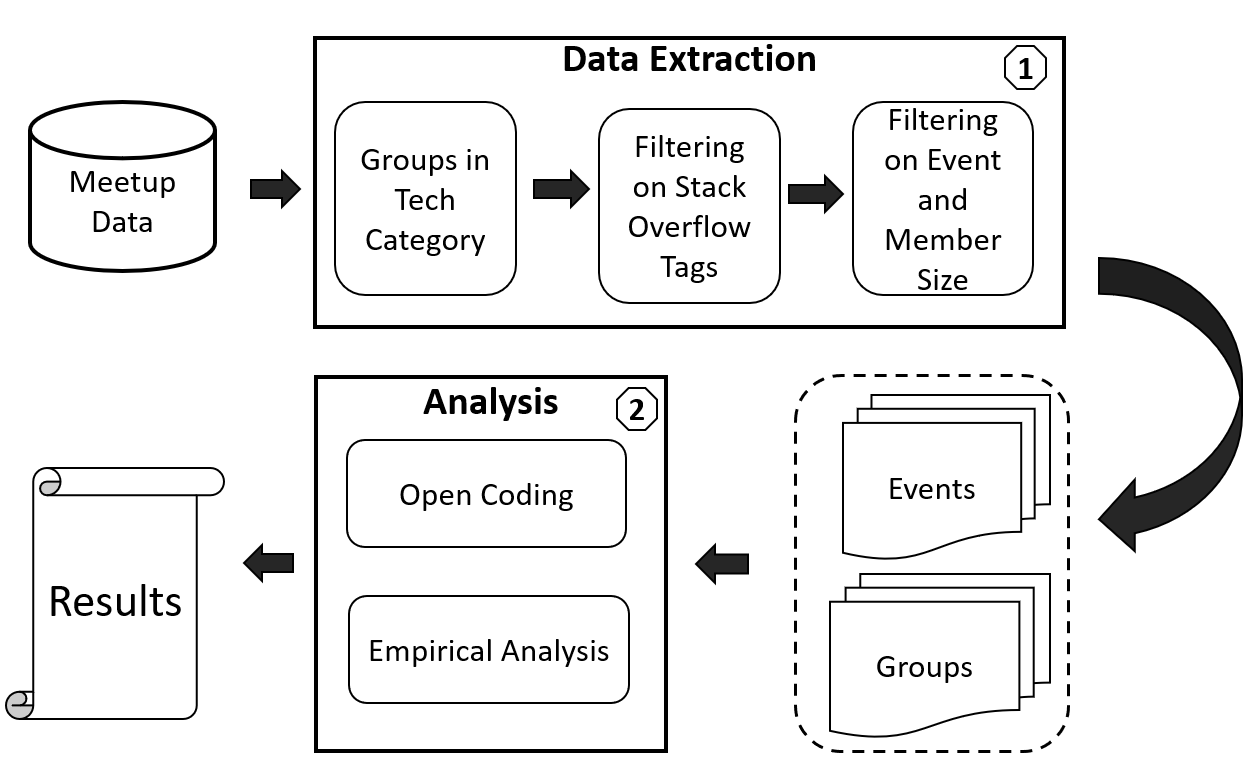}
	\end{center}
	\caption{Overall Process}
	\vspace{-0.5cm}
	\label{fig:system}
\end{figure}

In this section, we present our research setting. 
The overall process that we follow in our empirical study is illustrated in Figure~\ref{fig:system}. First, we extracted the data of any group related to software development using Meetup API and heuristics leveraging categories and topics in Meetup and tags in Stack Overflow. From the extracted data we took a random sample of the  events organized  by them, and then analyzed the same using open card sort~\cite{hudson2013card,spencer2009card} and subsequent manual labeling~\cite{aniche2018modern,uddin2017mining}. Then we answered a few research questions based on the empirical analysis of the extracted as well as coded data. Our main contribution and research questions are briefly described below.

To the best of our knowledge this empirical study is the first of its kind to analyze in depth software related offline social interactions (in our case Meetup related software events). We have also to build a topic association graph based on the data extracted which helped us to find and highlight inter relationships between various topics.




\subsection{Research Questions}
\label{subsec.rqs}
\vspace{-0.05cm}
\subsubsection{RQ1. What are the categories of events organized by Meetup groups related to software  development?}  Meetup events are known to help developers in keeping up-to-date~\cite{storey2017social}. However, to the best of our knowledge no study has yet explored what are the kinds of events held in Meetup groups related to software development. In this research question we have employed manual qualitative analysis to develop some categories of events organized by Meetup groups related to software development. Finding such categories gives some insights into the types of knowledge that is discussed in Meetup groups. 
\vspace{-0.2cm}
\subsubsection {RQ2. How popular is each event category?}  We also analyzed the popularity of events based on the categories developed in RQ1. This gives an insight into what categories are more popular among the software development community on Meetup. Building upon the initial insights gathered, in future a detailed study can be done to explore what factors determine the popularity of a Meetup group. This can help Meetup organizers to effectively organize events and make their groups popular.

\vspace{-0.2cm}
\subsubsection{RQ3. How diverse are Meetup groups with respect to gender?}
By analyzing the gender diversity of Meetup groups related to software development, we can observe if the results match to diversity ratios in other social media channels. Based on the insights found, future studies can focus on identifying the reasons for difference in diversity if found.

\vspace{-0.2cm}
\subsubsection{RQ4. How does the popularity of topics of Meetup groups change over	time?} 
By analyzing the change of the popularity of topics over time, we can observe which topic is rising or falling. This analysis can help developers to monitor whether the topics they know are still in demand.  It also helps developers to see if a new framework or library can be used to develop applications in future. 
%


\vspace{-0.2cm}
\subsubsection{RQ5. Can we observe any topic associations among group topics?}
By analyzing topics of Meetup groups related to software development, we can identify associations between technologies, suggesting the technologies that developers may want to study or use together.



\vspace{-.2cm}
\subsection{Data Extraction}
\vspace{-.15cm}
 

For our study we considered the data of groups on Meetup which are categorized under {\em Tech}\footnote{\url{https://www.meetup.com/topics/}} category. This category is assigned to a Meetup group by its organizers at the time of group creation. We made  use of an open source python client\footnote{\url{http://meetup-api.readthedocs.io/en/latest/index.html}}  based on Meetup's  RESTful API\footnote{\url{https://www.meetup.com/meetup_api/}} to collect the data in our work, using which we were able to extract 58,460 groups categorized under  {\em Tech} category. The data was extracted in the time period 16.04.2018-19.04.2018.

 Next, in order to improve the quality of our data set we applied a heuristic based filtering based on Stack Overflow \textit{tags}\footnote{\url{https://stackoverflow.com/tags}}. These \textit{tags} are used to describe the topics of questions asked on Stack Overflow, and  each tag generally represents a software engineering concept. We were able to get a list of 51,670 \textit{tags} from Stack Overflow archive\footnote{\url{https://archive.org/details/stackexchange}} on 21.04.2018. 
This  list is referred to as \textit{SOTagList} further in the paper unless stated otherwise. Most of the \textit{tags} present in the \textit{SOTagList} are single words, e.g.,  python, javascript, java. 
Some tags are composed of multiple words connected by a hyphen (as tags cannot contain space)\footnote{\url{https://stackoverflow.com/help/tagging}}, e.g.,  visual-studio, apache-spark, ruby-on-rails. In order to filter groups based on \textit{SOTagList} we converted all the \textit{topics} associated with each of the 58,460 groups,  as well as {\em tags} present in  \textit{SOTagList} to lower case. Then, we considered only those groups for further processing  for which 
	\begin{itemize}
		\item \textit{SOTagList} contained any word appearing as a \textit{topic} of a group, e.g., \textit{topic} ``python'' was present in \textit{SOTagList}
		\item \textit{SOTagList} contained the  hyphen separated form of words associated as a \textit{topic}  of a group,  e.g., \textit{topic} ``big data'' when converted to ``big-data was present in \textit{SOTagList} 		
		\item \textit{SOTagList} contained any word present in the word sequence associated as a \textit{topic}  of a group,  e.g., ``database'' in \textit{topic} ``database professionals''  is present in \textit{SOTagList} 
	
	\end{itemize}
		

After applying the above heuristic we were left with 56,175 groups. As many of the groups may be very small, we further filtered them by considering only those  groups which have at least $N$ members and have organized at least $M$ events. We chose the value of $N$ and $M$ to be 10 for our study. We also excluded the groups whose event and member or data was not publicly visible. After applying this level of filtering, we were left with  17,727 groups. On observing these group descriptions we found out that some of them are not related to software development such as \textit{The Vancouver Blogger Meetup Group}. To address this issue, we further applied one more level of filtering to keep only those groups which also contained a ``Software Development'' tag among their topics. We were left with 6,327 groups after applying this final level of filtering.


From these 6,327 groups we were able to extract 320,807 events in total. There were some events which repeat after certain intervals of times.  As the instances of such periodic events have same description and theme, we remove all but one instance. 
After removing the repeated occurrences of such events we got a total of  213,477 events.  Then we made use of as python package \textit{langdetetct}\footnote{\url{https://pypi.python.org/pypi/langdetect?}} to remove those events whose description language was not \textit{English}. In the end, we were left with 185,758 events. We then manually analyzed and labeled 452\footnote{This corresponds to a 95\% confidence level with 5\% error margin}  events from these 185,758 events into categories which were developed through open coding and subsequent manual labeling. The coding and labeling process will be discussed in detail in Section~\ref{subsec.opencoding}.

We also extracted member information from 6,308 groups. For 19 groups the member extraction failed as there was no group organizer at the time when the query for member information was made. We were able to extract information 3,123,498 unique members. Among these there were 2,610 members who were not active. We discarded such members and in the end we were left with information of 3,120,888 members.

\label{subsec.datapreprocessing}


\vspace{-.15cm}
\subsection{Content Analysis}
\vspace{-0.15cm}
In order to find the categories of events organized by Meetup groups related to software development we used  the open coding procedure~\cite{saldana2015coding}, a method of qualitative analysis of data  based on grounded theory methodology~\cite{corbin1990grounded}.  In particular we used the  open card sort~\cite{hudson2013card,spencer2009card} process which is used to develop categories from unorganized data.  We performed the card sort procedure on the sample data of 100 events sampled from the larger dataset described earlier in Section~\ref{subsec.datapreprocessing}.  
The technique used in our work is similar to what has been used in  various previous studies such as ~\cite{latoza2006maintaining,lo2015practitioners,sharma2015s,zimmermann2016card}. We first generated a card based on the description of each event. Each card contained the event description,  event id, event date, event group, and event's URL link. Then, each card was read and the  event description along with other details was discussed and iteratively sorted into categories or groups. In the first iteration, a code was assigned to an event description, and in all the subsequent iterations the codes assigned in previous iterations were analyzed to create higher level concepts or categories. 
For some event descriptions we were not able to merge them with any other categories, so we merged into a special category \textit{Others}. In the end, we were able to generate 9 categories including \textit{Others}. The  first and the last author of the paper together performed the  open card sort process. 

\begin{table}[h]
	\vspace{-0.2cm}
	\centering
	\caption{Progression of Agreements while Labeling}
	\vspace{-0.2cm}
	\label{tab.agreement}
	\begin{tabular}{p{1.2cm}|p{1.4cm}|p{1cm}|p{2.8cm}}
		
		\hline
		\textbf{Iteration } & \textbf{Absolute Agreement} & \textbf{Cohen's Kappa} &\textbf{Interpretation}\\ \hline
		1               & 0.800                & 0.700      &  Substantial     \\
		2               & 0.800                & 0.705      &  Substantial\\
		3               & 0.800               & 0.738       &  Substantial\\
		4              & 0.867              & 0.832         &  Almost Perfect\\ \hline
	\end{tabular}
\end{table}

 In order to increase our sample size, we further sampled 400 more events. During open coding earlier we had already come up with a coding schema. These 400 events were then coded by the first and second author based on the categories developed earlier.  The first author discussed the schema with the second author to arrive at a common understanding and clear any confusions. Then both authors separately labeled 30 events and then met together to discuss and further refine the coding schema if required. The authors continued the iterative process of independently coding and discussing afterwards until they were consistent in labeling. After 4 iterations were completed  both the authors were already achieving  substantial to perfect agreement, reaching a Cohen's Kappa~\cite{cohen1960coefficient} agreement score greater than 0.7 on all the iterations. After this iteration the rest of the data was split into two sets which were independently coded by first and second authors. The process is similar to what has been followed in~\cite{aniche2018modern,uddin2017mining}.

 During iterative discussions we came across some event descriptions where very little information was available on what  the type of event is. Also, for some events the descriptions were primarily in language other than English which made it hard to determine the type of events. For some other descriptions it was hard for labelers to assign a  single category to the event based on the event description. For all the aforementioned 3 cases the labelers could assign them to an ``Unsure" set. After both coders finished the labeling we had in total 52 events which were put in the ``Unsure" set. We dropped such events from our final dataset. The final data set is of size 452 events (and not 448 as we had to split 4 events whose descriptions  contained two types of events). Note that ``Unsure" set is different from the ``Others'' category where we were able to identify what the event is about but not able to merge it with other categories.


\label{subsec.opencoding}

\vspace{-0.15cm}
\subsection{Data Characteristics}
\begin{table*}[]
		\centering
\caption{Descriptive Statistics of Our Dataset}
\vspace{-0.25cm}
\label{tab.datastats}
\begin{tabular}{@{}l|l|l|l|l|l|l|l|l|l@{}}
	\hline
	&          &\textbf{ Count}   & \textbf{Mean}     & \textbf{Std Dev}      & \textbf{Min}     & \textbf{Q1}      & \textbf{Median}  & \textbf{Q3}      & \textbf{Max}             \\ 
	\hline
	\multirow{3}{*}{\textbf{Groups}} & Members  & 6,327& 1,075.94  & 1,479.65  & 12  & 303.50  & 629 & 1,285.50 & 23,043        \\\cline{2-2}
														& Rating   & 6,327 & 4.53     & 0.98     & 0.00    & 4.60    & 4.75    & 4.88    & 5.00            \\ \cline{2-2}
														& Topics   & 3,282    & 22       & 173      & 1       & 1       & 2       & 6       & 6,327            \\ \cline{2-2}
	\hline
		\multirow{2}{*}{\textbf{Events}}  & Duration& 100,573  & 5:25:13  & 15:53:46 & 0:00:00 & 02:00:00 & 2:15:00 & 3:00:00 & 14 days \\ \cline{2-2}
		
	& Rating   & 185,755  & 2.63 & 2.37 & -1      & 0       & 4       & 5       & 5               \\ \cline{2-2}
	\hline
	\textbf{Members}                        & Topics   & 2,370,094 & 18       & 15       & 1       & 6       & 14      & 28      & 67              \\  \cline{2-2}
	\hline
\end{tabular}
\vspace{-0.25cm}
\end{table*}




Table~\ref{tab.datastats} shows some descriptive statistics related to the data used in our work.  For \textit{Groups}, the average number of members per group is 1,076, however the median number of members is lower at 629. There are some groups with  very high membership count such as {\em DC Tech Meetup}  (\url{https://goo.gl/9VG5qc}) with 23,043 members.  We also looked at the mean and median ratings assigned to the \textit{groups} and there seems to be less deviation in the ratings assigned to groups, with mean and median being close to each other, 4.53 and 4.73 respectively.

 For \textit{Events}, looking at their time duration we found that the mean average duration of each event is 5 hours and 25 minutes, whereas the median duration is quite low at 2 hours and 15 minutes. There are some events which last over few days, such as a 14 day event related to {\em Ruby Workshop} (\url{https://goo.gl/tJoH9w}) by the group {\em Girl Develop It Ann Arbor}. For events, the ratings seem to be quite divergent with a mean rating score of 2.63 and a median rating score of 4. 

Table~\ref{tab.datastats} also shows a summary of topics associated with groups and members. The mean number of topics associated with groups are 22 per group, which is slightly higher than the 18 topics per member. However, when considering the median scores, most groups only have 4 topics associated with them as compared to 14 topics associated with members. 

\begin{figure}[!htb]
	\vspace{-0.2cm}
	\begin{center}
		\includegraphics[width=3.2in,height=1.7in]{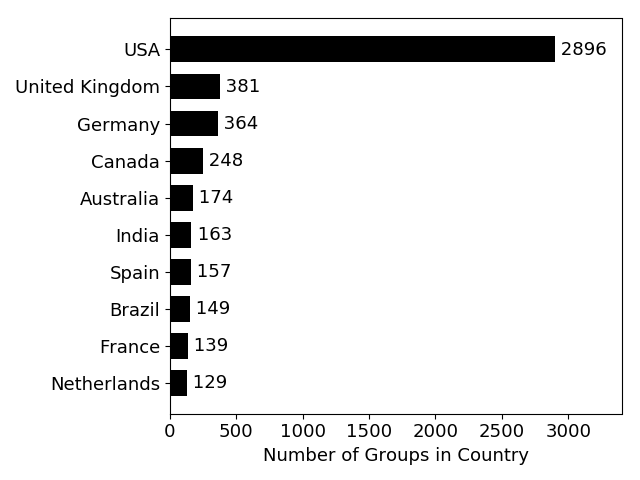}

	\end{center}
	\vspace{-0.3cm}
	\caption{Top-10 Countries by Numbers of Groups}
	\vspace{-0.2cm}
	\label{fig:group_country}
\end{figure}

\begin{figure}[!htb]
	\vspace{-0.2cm}
	\begin{center}
		\includegraphics[width=3.2in,height=1.7in]{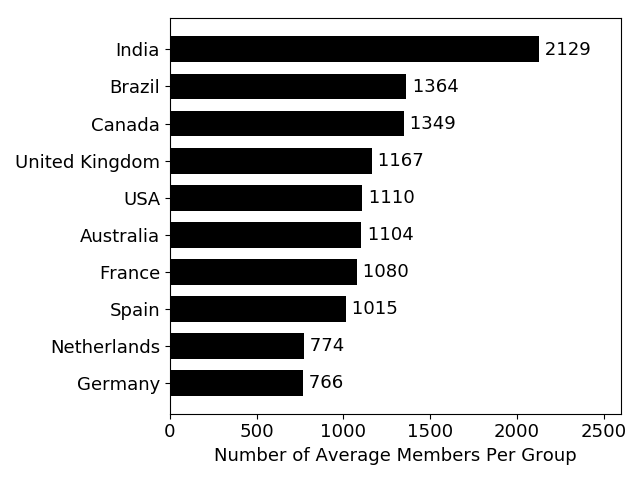}
	\end{center}
		\vspace{-0.4cm}
	\caption{Members Per Group in Countries having most Groups (Top-10)}
	\vspace{-0.2cm}
	\label{fig:group_country_members}
\end{figure}

We were able to extract a total of 6,327 Meetup groups as described earlier in Section~\ref{subsec.datapreprocessing}. 
From a geographical perspective, we looked at how 6,327 groups are spread across different regions. We found that the groups in our dataset span across 107 countries.  Figure~\ref{fig:group_country} shows the number of Meetup groups for the top-10 countries containing the most groups.
All the countries in this list except India and Brazil are developed nations (based on Human Development Index\footnote{\url{https://en.wikipedia.org/wiki/List_of_countries_by_Human_Development_Index}}). Interestingly, when we look at the average number of members per  group in these top-10 countries having most groups, India and Brazil occupy the top 2 spots respectively as can be seen in Table~\ref{fig:group_country_members}.




\label{subsec.eda}




%
%
%
%
%
%
%

%
%
%
%
%
%
%
%

\vspace{-0.2cm}
\section{Results}
\label{sec.results}
In this section, we present the results of our analysis conducted in this  study.  Our work aims to answer the 5 research questions mentioned in Section~\ref{sec.intro}




\vspace{-.15cm}
\subsection{RQ1. What are the categories of events organized by Meetup groups related to software  development?}
\label{sec:RQ1}
In this research question we explored  if the events organized by Meetup groups related to software development can be grouped into some meaningful categories. 
We were able to determine 9 categories of events (8 main categories +``Others'') using the open coding methodology~\cite{saldana2015coding} described earlier in  Section~\ref*{subsec.opencoding}. The categories found are described below along with a relevant event as an example.

 \textbf{Talks by Domain Experts:}  200 of the event descriptions analyzed were related to an event where talks or presentations were given by some domain experts.  The domain experts consisted of mostly developers, technical managers, entrepreneurs and CEOs.  Most of the talks are technical in nature where an experienced developer explains or introduces a core software  concept. 
 Sometimes multiple experts came together and participated in panel discussions. In other talks technology management principles such as SCRUM were presented. Other talks especially by entrepreneurs and CEOs focused on a technical product or feature being developed by their companies.  An example of an event in this category 
 is shown below: 
 

 \begin{framed}
	\vspace{-0.25cm}
	\noindent \textit{Meetup group}  : Boston-Predictive-Analytics \\
	\noindent \textit{Description} :  ... Rani Nelken of Outbrain (\url{http://www.outbrain.com/}) has graciously offered to present on Bayesian Classification....
	\\
	\noindent \textit{Event URL:} \url{https://tinyurl.com/y2ncfue5}
	\vspace{-0.25cm}
\end{framed}


\textbf{Hands-on Sessions:}  This category contains events where a domain expert does not only give a talk or presentation but is also involved in actively guiding and helping other participants to perform some hands-on tasks or activities related to a topic of presentation. Such kind of events were many a times marked with a request for the participants to bring their own laptops so as they can practice the exercises that  follow a talk. The sessions organized in such events can range from introductory  to advanced. 
Many a times such events required the participants to pay a fee. An example of such an event is mentioned below:

%


 \begin{framed}  
 	\vspace{-0.25cm}
 	\noindent \textit{Meetup group} :  WaikatoLinuxUsersGroup  \\
 	\noindent \textit{Description} : ... This is a GNU/Linux-focused workshop where people can bring their PCs, Laptops, Pi's, Android devices etc for trouble-shooting and to learn or try out new skills ... 
 	\\
 	\noindent \textit{Event URL} : \url{https://goo.gl/5PLL7P}
 	\vspace{-0.25cm}
 \end{framed}

\textbf{Conferences:} The events in this category were created mostly to notify the group members of any upcoming conferences. Different from talks and hands-on sessions, conferences were longer and bigger events that often span multiple days, with participation by many speakers, and covered a more diverse range of topics. Sometimes it also involved call for proposal, participation, or volunteering. An example event is shown below:  

\begin{framed}  
	\vspace{-0.25cm}
	\noindent \textit{Meetup group} :  jsmeetup  \\
	\noindent \textit{Description} : Call for Speakers HTML5DevConf continues to grow as the largest JavaScript and HTML5 conference  ...
	\\
	\noindent \textit{Event URL} : \url{ https://goo.gl/ZT8mwM}
	\vspace{-0.25cm}
\end{framed}

%

\textbf{Open Discussions :} The events in this category did not have any predefined agenda  and no speakers were scheduled to speak in advance. Most of these were open house sessions where any of the participants could speak on any topic loosely related to the topics associated with groups. It included events such as round table discussions, impromptu experience sharing  sessions, study groups, code jams, etc. An event categorized into this category is shown below:
 \begin{framed}  
 	 	
	\noindent \textit{Meetup group} :  london-software-craftsmanship  \\
	\noindent \textit{Description} : 
	 ... Do you want to discuss an approach pattern or technology and see what others think? Or perhaps discuss a design challenge youre facing? Come along to the Software Craftsmanship round table ...  \\
	\noindent \textit{Event URL} : \url{https://goo.gl/3mFmwQ}
	 
\end{framed}

\textbf{Social Events:}  This category includes core networking events where participants  were invited to social dining and/or drinking sessions where they could interact with other invitees. Such events include kick-off parties, award ceremonies, etc. An example event is shown below:
									 

 \begin{framed}  
 	\vspace{-0.25cm}
 	\noindent \textit{Meetup group} :  Windy-City-Tech-Meetup \\
 	\noindent \textit{Description} :  Talk with others in Tech, Big Data, Business Intelligence, Open Analytics, etc., and meet new business contacts at River North's Trophy Room on December 28th, while sipping sponsored cocktails and beer.
 	\\
 	\noindent \textit{Event URL} : \url{https://goo.gl/5nBLa8}
 	\vspace{-0.25cm}
 \end{framed}

 	\vspace{-0.05 cm}
 	\textbf{Competitions:}  In the events under this category the participants usually formed teams, and then competed with one other on certain technical tasks. The most common type of event under this category was \textit{Hackathon} where teams had to come up with a usable software product in some days or hours. Sometimes the events were held specifically by a technology product company where they offer rewards to the participants in order to find bugs in their products. An example event is shown below:
 	

 	\begin{framed}  
 		\vspace{-0.25cm}
 		\noindent \textit{Meetup group} : San-Francisco-Hackathons  \\
 		\noindent \textit{Description} : ... The Hackathon will consist of 8 teams with up to 6 members. You may register as an individual or bring an entire team. If you register as an individual, we will find a team for you ...  \\
 		\noindent \textit{Event URL} : \url{https://goo.gl/SCoHUz}
 		\vspace{-0.25cm}
 	\end{framed}

\vspace{-0.05 cm}
\textbf{Administrative Events:} These events were primarily organized  to discuss among the  group organizers and volunteers the roles and responsibilities each group member would take. Sometimes other organizational aspects such as what kind of events to organize in future were also discussed in such events. 
 The event shown below is one of the events that has been assigned into this category:


\begin{framed}  
	 	\vspace{-0.25cm}
	\noindent \textit{Meetup group} :  Evansville-Technology-Group \\
	\noindent \textit{Description} : ... Come join a round table discussion on technology in Evansville and help us plan our meetups for the year. We need your input to make sure we are providing the topics and events that everyone is interested in ....   \\
	\noindent \textit{Event URL} :  \url{https://goo.gl/qRhPHw}
	 	\vspace{-0.25cm}
\end{framed}

\vspace{-0.05 cm}
\textbf{Job Fairs:} These are the events which bring together recruiters and job seekers; the focus being on software related jobs. The following event is an example:
  

 \begin{framed}  
    	\vspace{-0.2cm}
    	\noindent \textit{Meetup group} :  Girl-Develop-It-Boulder-Denver  \\
    	\noindent \textit{Description} : .... The Tech Jobs Tour is coming to Denver and looking to connect with techies and community-focused individuals in the city ...\\ 
    	\noindent \textit{Event URL} : \url{https://goo.gl/qr8zbe}
    		\vspace{-0.2cm}
 \end{framed}

 \vspace{-0.1 cm}
\textbf{Others:} Few events in our sample could not be categorized into any of the 8 above-mentioned categories. Since the remaining events are different from one another, we put all of them in a broad category {\em Others}. Some of the events  moved into  {\em Others} were study tours, marketing events, etc.

\vspace{-.15cm}
\subsection{RQ2: How popular is each event category?}
\label{sec.RQ2}
\begin{figure}[!htb]
	\begin{center}
			\vspace{-0.25cm}
		\includegraphics[width=3.2in,height=2in]{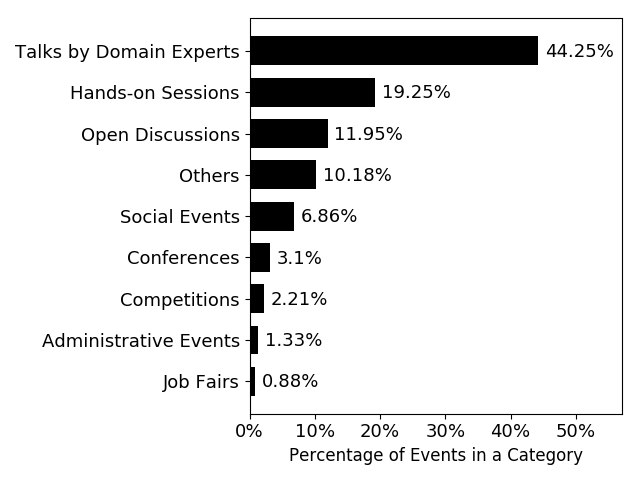}

	\end{center}
	\vspace{-0.35cm}
	\caption{Popularity of Categories in Sampled Event Data}
	\label{fig:RQ2}
	\vspace{-0.35cm}
\end{figure}

In this research question we analyzed the popularity of each event category in our labeled data. We define popularity based on how many events are included in a category. To do this we count the number of events categorized into each category and then plot a bar graph of percentage of events occurring in each category. Figure~\ref{fig:RQ2} shows a bar graph showing the popularity of each category. 
The percentage calculation was done on 452 total events. We can observe from the graph that  {\em Talks by Domain Experts} is the most frequent category of events organized by Meetup groups, followed by events related to {\em Hands-on Sessions} and {\em Open Discussions}.


We also did a popularity analysis based on the number of people interested in an event.  
For most events, a field called \textit{yes\_rsvp\_count} is present, which specifies the total number of people who confirmed participation for an event. We used \textit{yes\_rsvp\_count} as a proxy for estimating how many people are interested in the event. There is another field known as \textit{rsvp\_limit} which specifies the total number of people allowed for the event. We only considered those events for which values of both these  fields were present in our dataset. We were able to find 162 events spread across all 9 categories which had both the fields.  For each event, 
we calculate a metric called $Event\_Attention$ by dividing the value of \textit{yes\_rsvp\_count} by value of \textit{rsvp\_limit} and then averaging it over all the events for the respective category. 

\begin{figure}[htpb]
	\begin{center}
		\vspace{-0.35cm}
	\includegraphics[width=3.2in,height=2in]{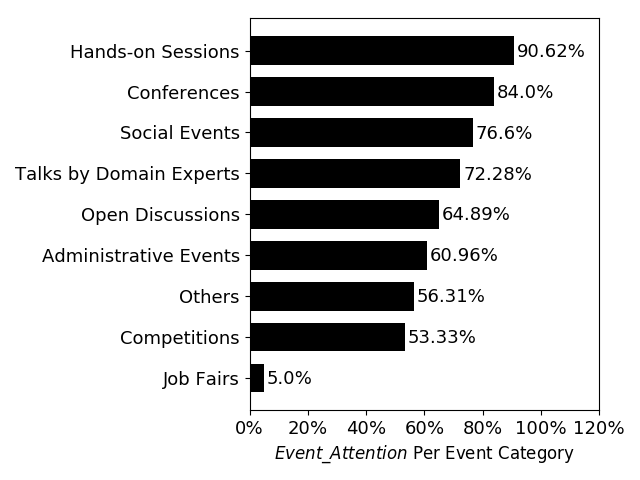}
	\end{center}
	\vspace{-0.35cm}
	\caption{Popularity of Event Categories based on Interest}
	\label{fig:RQ2b}
	\vspace{-0.35cm}
\end{figure}

Figure~\ref{fig:RQ2b} shows a bar graph showing the popularity of each category of events when the popularity is calculated based on  $Event\_Attention$. The category  {\em  Hands-on Sessions} is the most popular category with the highest $Event\_Attention$ of 90.62\%. In {\em Hands-on Sessions}, the participants are generally required to bring their own laptops so that they can work on  various tasks and exercises discussed in the event and also ask other people for help. The more interactive nature of such events seems to result in high participation in such events. The participants seem to be less interested in {{\em Competitions} as it may be difficult to form teams. Another interesting observation is that events related to the category {\em Job Fairs} have low values for $Event\_Attention$, which may be an indicator that software developers prefer other channels of contacting recruiters.

\vspace{-.15cm}
\subsection{RQ3: How diverse are Meetup groups with respect to gender? }
\label{sec:RQ3}
\begin{table}[]
	\caption{Descriptive Statistics of Gender Distribution}	
	\label{tab.genderstats}
	
	\begin{tabular}{@{}llllll@{}}
		\toprule
		& 			\textbf{Groups} & \textbf{Mean} & \textbf{Median} & \textbf{Max} & \textbf{Deviation} \\ \midrule
		\textbf{Male}      & 6196           & 600.91         & 351.50             & 12985        & 824.04            \\
		\textbf{Male \%}   & 6196           & 58.25         & 60.57           & 92           & 14.42              \\
		\textbf{Female}    & 6196           & 226.67           & 107             & 9369         & 412.22                \\
		\textbf{Female \%} & 6196           & 19.82         & 17.51           & 91.78        & 11.70              \\ \bottomrule
	\end{tabular}

\end{table}

We also analyzed how different genders are represented among various groups. As discussed in Section~\ref{subsec.datapreprocessing} we had extracted information 3,120,888 members. For these 3,120,888 members, we tried to resolve their gender using the name and country information. For the purpose of resolution, we used the name and country information of members as input to} \textit{genderComputer}~\cite{vasilescu2013gender,vasilescu2015gender}. We were able to determine gender of 2,618,848 members which constitutes a fraction of 83.91\%. For other cases, either the gender determination failed or it could not be determined if the person is male or female. Based on this gender determination process we were able to find at least 1 female or 1 male member for 6,196 groups. For other groups, we could not successfully resolve gender for any of their members. We then computed the percentage of male and female members based on the total members count present in group which was extracted earlier as described in Section~\ref{subsec.datapreprocessing}.  The total members count includes members whose gender could not be determined.


Table~\ref{tab.genderstats} shows some descriptive statistics with respect to gender distribution across the groups. We can see that 19.82\% of the members on an average are female. This numbers is larger as compared to results reported in previous studies, 3-9\% on Github~\cite{vasilescu2015gender,david2008community} and 7\% on Stack Overflow~\cite{vasilescu2013gender}. This insight shows that females may be more comfortable to participate in events in Meetup; the reasons for  which deserves more research and understanding in future studies. 

For each of the 6,196 groups there was 1 organizer associated per group, so we also did an analysis of the gender diversity among organizers. For 5,008 groups we were able to determine the gender of the organizer of the group and found that for 853 groups (i.e 17.03\%) the organizers were female. 
Among these 5,008 groups we found that for the groups where the organizer was female, the mean of \textit{female percentage} for such groups is 28.97\%, as compared to 17.60\% for the groups where the organizer was male. To validate that \textit{female percentage} for groups with female organizers is indeed significantly different from the groups with male organizers, we performed the Mann-Whitney U test~\cite{mann1947test} on distributions of \textit{female percentage} for both cases. The test gave a p-value which is less than 0.05, and thus we can say that the distributions are significantly different. We also computed the Cliffs Delta~\cite{cliff1993dominance} statistic for the two distributions and found the deta value to be 0.413 (medium). This suggests that if the organizer of a Meetup group is female, they tend to attract more females to the group.





\vspace{-.15cm}
\subsection{RQ4: How does the popularity of topics of Meetup groups change over time?}
\label{sec:RQ4}

In this research question, we analyzed what topics are more popular with Meetup groups in recent times. As explained in Section~\ref{sec.background}, whenever a group is created by a member, she/he can associate some fine grained topics with groups represented by keywords such as ``python", ``java", ``sql", etc.  We analyze the trends of topics over time. To do this we define impact of topic $t$ in year $y$ as follows. 

\begin{figure*}[h]
	\vspace{-0.5cm}
	\subfigure[Top 10 topics with increasing popularity]
	{
		\includegraphics[width=3.5in,height=2.7in]
		{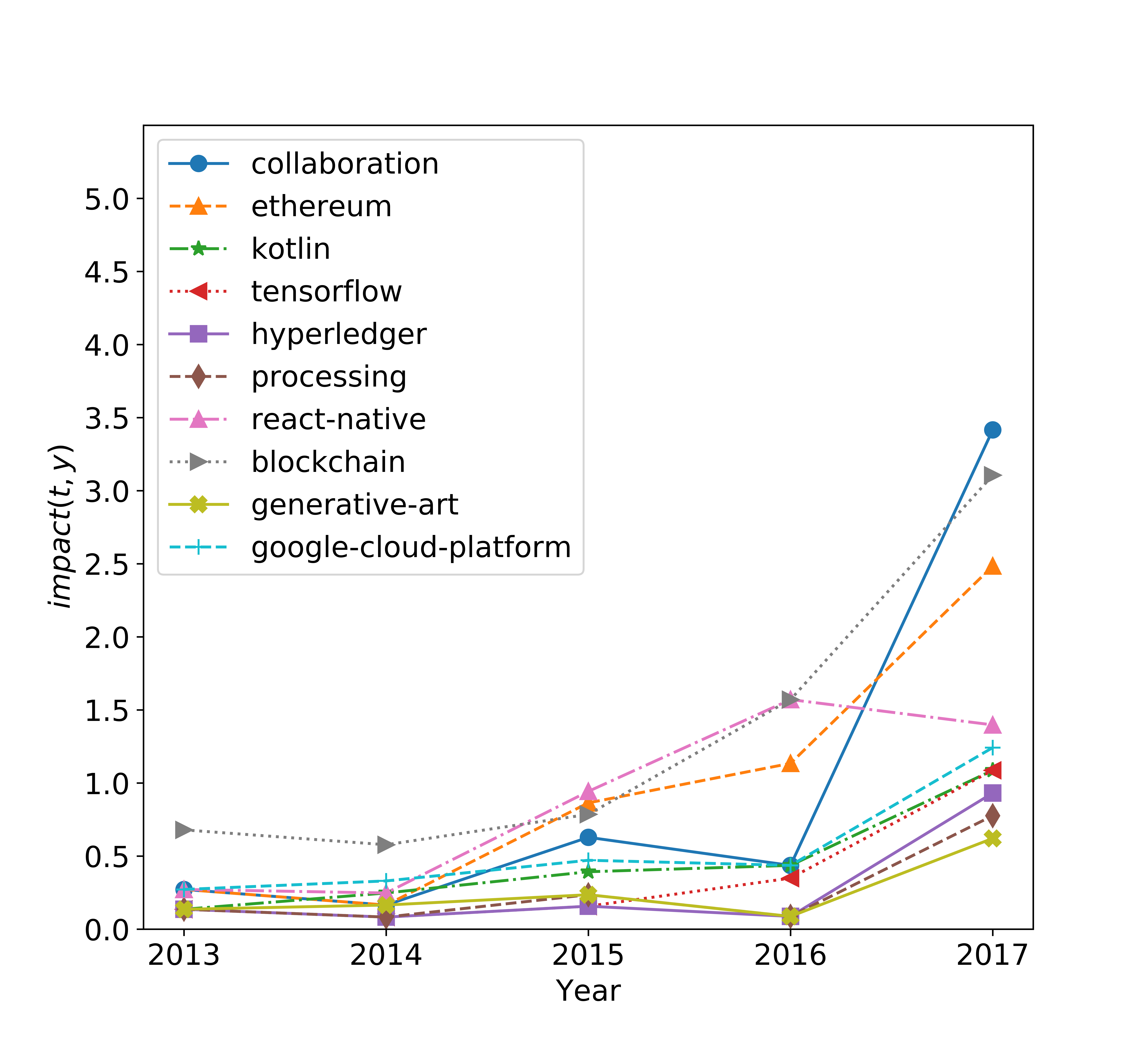}
		\label{fig:RQ3a}
	}
	\hspace{-0.5cm}
	\subfigure[Top 10 topics with decreasing popularity]
	{
		\includegraphics[width=3.5in,height=2.7in]
		{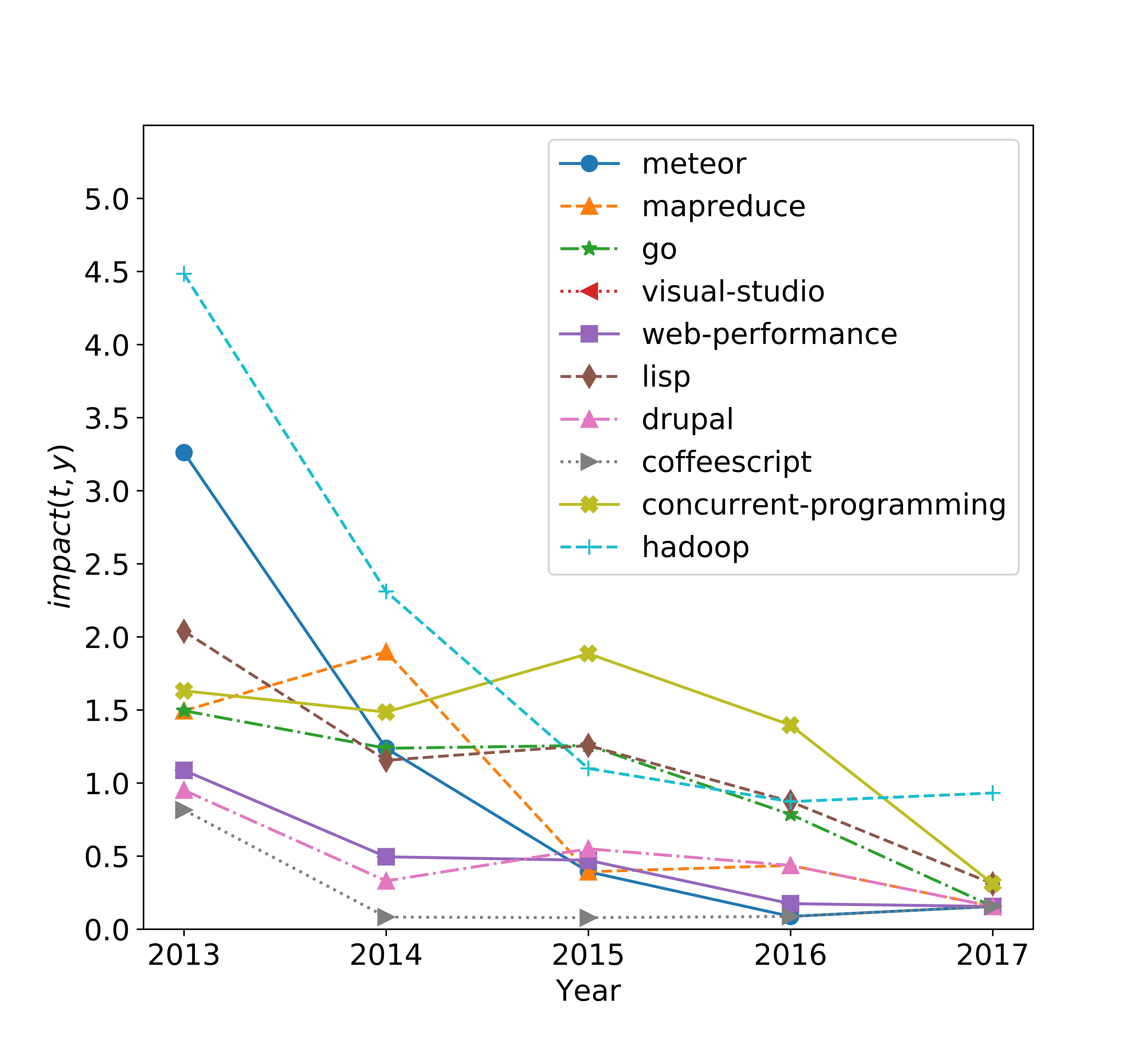}
		\label{fig:RQ3b}
	}	
	
	\caption{Change in popularity of topics}
	\label{fig:RQ3}
\end{figure*}



\vspace{-0.25cm}
\[
impact(t, y) =  \frac{G(t,y)+1}{\sum_{t \in T} {G(t,y))} +1}
\]
\vspace{-0.25cm}

\noindent where $G(t,y)$ is the set of all groups created in year $y$ associated with a topic $t \in T$,  where $T$ is the set of all topics found across all groups. 1 has been added to both numerator and denominator to avoid  0 value for $impact(t,y)$. We were able to find 3,282 topics across the 6,327 groups considered in our study. Many of the  topics  in this list are quite generic such as ``computer programming". So we further narrowed down the list of topics by considering only those topics which were also present on \textit{SOTagList}, which consists of all tags in Stack Overflow (described in detail in Section~\ref{sec.setting}).  After this filtering step, we were left with 729 topics which constituted set of topics $T$, for which growth of associated Meetup groups was analyzed.


In this work, we focused on topics that are popular in the recent past, so we analyzed the impact of topics over a 5-year period from 01.01.2013-31.12.2017. This 5-year period constitutes the set $Y$. For each year $y \in Y$, the impact for a topic $t$ for  that year $y$ was calculated as $impact(t, y)$ which has been defined earlier. Next, we calculated the value of $Growth(t)$ for all 729 topics in the set $T$. $Growth(t)$ measures the percentage change in impact of a topic $t$ from year 2013 to 2017. It was calculated as follows
\vspace{-0.05cm}
\[
Growth(t) = \frac{ 100*[impact(t, 2017)-impact(t, 2013)]}{impact(t, 2013)}
\]
\vspace{-0.25cm}

Then we found out the top 10 topics with the highest values for $Growth(t)$ and also the bottom 5 topics having lowest $Growth(t)$ values. These have been plotted in Figure~\ref{fig:RQ3a} and  Figure~\ref{fig:RQ3b} respectively.

Figure~\ref{fig:RQ3a} shows a line graph showing the \textit{impact} of top 10 topics, which was determined based on their $impact(t, y)$ values. The most rapidly growing topic in Meetup groups related to software development is  $collaboration$. This signals that more developers are joining groups which promote collaborative activities. An example of such a group is \textit{Tokyo Python Society Club} (\url{https://goo.gl/grJqhC}), which describes itself as open to anyone who is interested to improve their knowledge or teach others. A look at the recent organized events shows that many of them can be classified as \textit{Hands-on Sessions}, a category developed as part of RQ1 in Section~\ref{sec:RQ1}. The other popular topics relate to some recent technologies such as {\em ethereum} and {\em hyperledger} are both related to another topic {\em blockchain}, which in very simple terms is a list of records linked and secured using cryptography. {\em Blockchain} itself is the $8^{th}$ most popular topic based on growth metric $Growth(t)$ defined earlier. Topic {\em ethereum} refers to a  blockchain based distributed computing platform, whereas {\em hyperledger} is a project based on open source blockchains. The rise in Meetup groups related to these technologies points to an increased interest in the software development community to get up-to-date with these platforms. An example of a Meetup group related to both these topics is {\em Blockchain Developers United} (\url{https://goo.gl/rdT23g}).  Most events organized by these group are related to \textit{Hands-on Sessions} and \textit{Talks by Domain Experts} categories defined earlier.
Some of the other popular topics shown in Figure~\ref{fig:RQ3a} are {\em kotlin}, {\em tensorflow}, {\em react-native}, and {\em google-cloud platform}. All of these are examples of libraries, frameworks or platforms used by software developers. The topic {\em kotlin} refers to a statically typed programming language that can be used for multi-platform application development and is thus becoming very popular lately. The topic {\em react-native} is a framework used to develop cross-platform applications in javascript. The topic {\em tensorflow} corresponds to an open-source software library which is very popular in developing machine learning based applications especially the ones based on neural networks.



From Figure~\ref{fig:RQ3b} we can see that the {\em hadoop} (a big data platform) is the topic whose popularity is falling most rapidly. This can be attributed to rise of competing platforms such as {\em spark, BigQuery} etc. The 
topics  such as {\em meteor} and {\em coffeescript}  which are javascript frameworks are also losing popularity, which may be attributed to  popularity  of another similar library {\em react-native}, which in fact is the $7^{th}$ most popular topic based on $Growth(t)$  metric. For the years 2017 and 2016, we did not observe any new Meetup group related to {\em meteor} topic. Also techniques such as {\em mapreduce} have become quite commonplace so the popularity of this topic also seems to be going down. A surprising presence can be seen here of the programming language {\em go} which despite being pushed by Google, seems to be still losing popularity. The other topics which exhibit a downward trend in terms of their impact are topics related to system performance: {\em web-performance} and {\em concurrent-programming}. As cloud based services have become quite streamlined and stable over past few years the focus on such topics is expected to go down. 

\vspace{-.1cm}
\subsection{RQ5:  Can we observe any topic associations among group topics?}
\label{sec:RQ5}
\begin{figure}[h]
	
	\centering
	\includegraphics[width=1\linewidth]{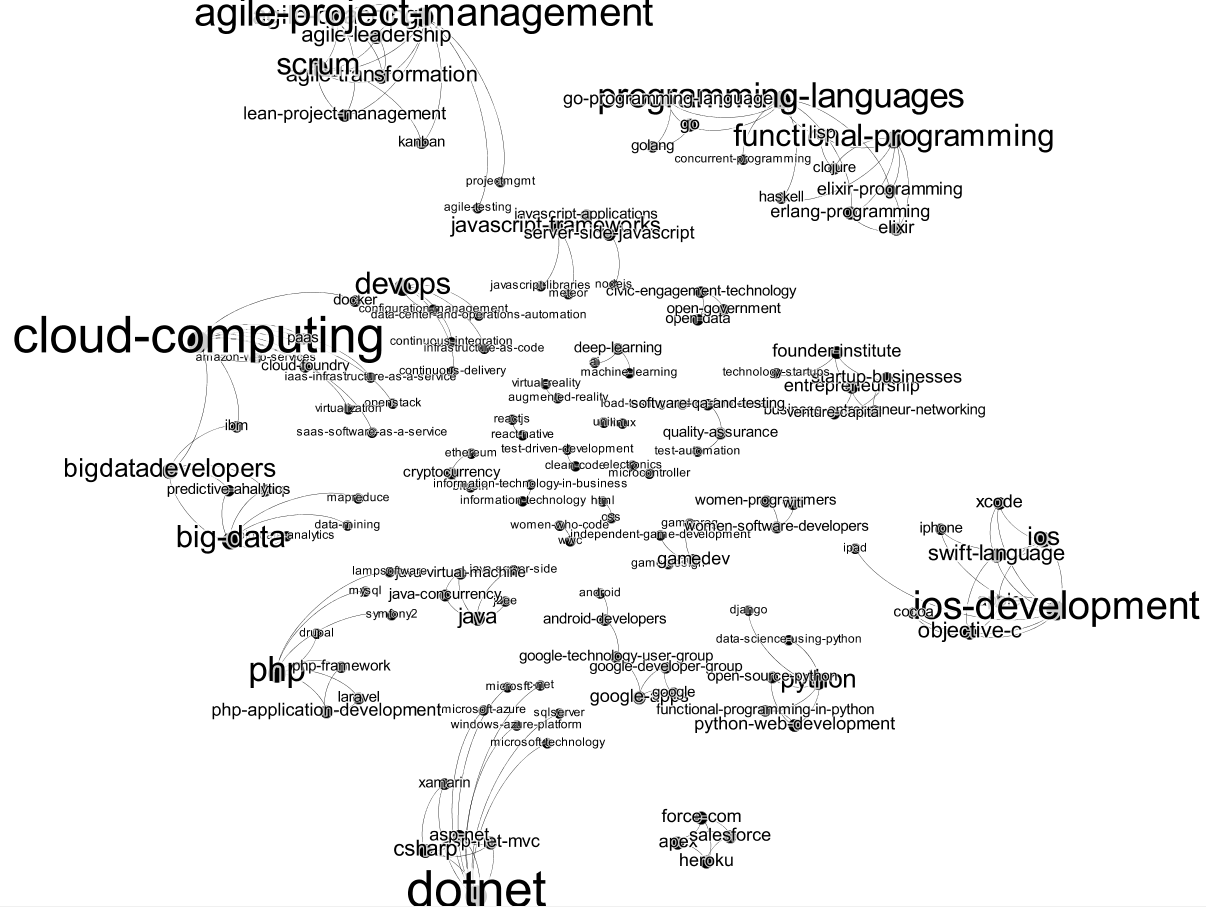}
	\vspace{-0.2cm}
	\caption{Network of topic co-occurrence}
	\label{fig:group-topic-cooccurrence}
	
\end{figure} 
\vspace{-0.15cm}

We have analyzed associations for topics related to groups in our dataset to see if we can find some interesting associations based on these topics. To discover network of topic co-occurrences, we applied association rule mining~\cite{agrawal1994fast} on the topics of the groups, with each group treated as a transaction and each topic as an item. We used a process similar to what has been done in~\cite{chen2016mining}. During this process, we first find the sets of topics whose ratio of co-occurrence compared to total number of groups (i.e. support) exceed certain minimum support value {\em minsup}. Subsequently we mine association rules in form of $topic_1 \implies topic_2$ that meets or exceeds certain confidence threshold {\em minconf}, where {\em confidence} being the percentage of groups having both $topic_1$ and $topic_2$ compared to groups having $topic_1$. For each rule, we also compute the {\em lift}, which is the ratio of confidence of the rule to the expected confidence given the dataset, and use it to filter the rules that are potentially most useful (i.e., having high lift). Afterwards, we generate a graph with each topic in the final set of rules being represented as a node, and rules $topic_1 \implies topic_2$ being represented as edge between $topic_1$ to $topic_2$. The result is shown in Figure~\ref{fig:group-topic-cooccurrence} at minimum support = 0.005, minimum confidence
= 0.75, and minimum lift = 5. The figure has also been shared at \url{http://tiny.cc/vdqdbz} for better visibility.


We observed that in addition to natural association between, for example, programming language and library/framework written in that language (e.g. {\em dotnet} with {\em csharp} and {\em asp-net-mvc}), there are topical clusters of concepts such as {\em agile, scrum},  etc. that are associated with management of software development. Some of the clusters contain topics such as  {\em women-programmers, woman-software-developers} which point to the effort the software community is making in addressing diversity issues by organizing events focused on certain under represented groups. We also observe clusters comprising  of a mix of specific technologies, such as {\em docker} with {\em devops} and {\em cloud-computing}. This highlights technologies that are closely relevant to one another; and one may want to acquire expertise in these technologies together.

\vspace{-.15cm}
\subsection{Threats to Validity}
\label{sec.threats}
Threats to \textit{ internal validity }relate to errors  that may have occurred during experiments and labeling of data. We checked our code multiple times, but there still may be errors that we  may have missed out. The labeling process involved 3 persons, 2 PhD students and 1 research engineer having more than  20 years of professional software development experience spread among them. 
We also computed the inter-rater agreement for the labeling task using the measure of Cohen's Kappa~\cite{cohen1960coefficient}. Threats to \textit{external validity} relate to how generalizable our findings are. We have tried to mitigate this threat by randomly sampling events. Also as seen in Table~\ref{tab.datastats} we can see that our event dataset spans across various topics.  Another threat relates to use of names for gender resolution, as users may be using aliases (and not their correct names) on Meetup groups. To address this we use \textit{genderComputer} which has been has shown high precision on the gender determination task~\cite{vasilescu2013gender,vasilescu2015gender}.

\vspace{-0.2cm}
\section{Discussion}
\label{sec.disccussion}
From our findings presented in Figure~\ref{fig:RQ2},  we can clearly see that the most popular category of events is \textit{Talks by Domain Experts}, followed by \textit{Hands-on Sessions} and { \em Open Discussions}. Also gender analysis on group members shows that female representation in the Meetup groups is higher as compared to other collaborative sites frequently used by developers. Based on the analysis of topic popularity, it can be observed that topics related to broad themes of {\em blockchain, machine learning, cross-platform development} etc. are becoming popular over last few years. Based on the above results we discuss below some implications of the result.

\subsection{Implication for Researchers}
We found that female representation in the Meetup groups being 19.82\% which is higher than what has been reported for other social channels,3-9\% on Github~\cite{vasilescu2015gender,david2008community} and 7\% on Stack Overflow~\cite{vasilescu2013gender}. Females tend to participate less if the technical barrier is too high~\cite{mendez2018open}. So further research is required to validate if the female participation is higher only for Meetup groups which have a low technical barrier or if it is standard across all the Meetup groups. We also found empirical evidence that average female participation in Meetup groups is higher if the group organizer is female. If the organizer of a group is female then the 
social barrier of participation decreases which has been cited as a reason for increased participation~\cite{lee2019floss}. However,further research is required to understand if there are any other factors which may be contributing to higher participation, and if those factors can be replicated elsewhere to improve female participation. 

 
We found that \textit{Talks by Domain Experts} is the most popular category and it highlights the importance software developers give to continual learning.  Also from Figure~\ref{fig:RQ2b}, it can be observed that the interest in the events related to category \textit{Hands-on Sessions}  is very high.  One reason for this phenomena is that in  \textit{Hands-on Sessions} participants actually work on exercises and modules, and thus have better understanding of the topic that is being discussed in the event. The popularity of category \textit{Hands-on Sessions} provides additional evidence to the value of the flipped-classroom setting in teaching programming classes
~\cite{kiat2014flipped,fulton2012upside}. Further research can be conducted on the data related to organization and participation information of popular event categories, which can then be used to understand what helps developers in their knowledge seeking experience. Such insights can be used by universities and other organizations involved in software engineering education, in order to improve the effectiveness of their programs and courses.

We found that topics related to broad themes of {\em blockchain, machine learning, cross-platform development} etc. are becoming popular over last few years. An implication for software engineering researchers is to pursue more research into the software engineering challenges related to above areas. There already has been some initial traction in these areas~\cite{tian2018deeptest,khomh2018software,semla19}. Another implication is opportunities to investigate the reasons for rise of cross-platform development technologies like {\em kotlin and react-native}. Given the high level of excitement among practitioners in these areas, more work is needed. Some work has already begun in this area~\cite{mateus2018empirical,mateus2019adoption}, but more is needed.


%
%
%
%
%
%
%
%

\subsection{Implication for Practitioners}

The empirical evidence that average female participation in groups is higher if the group organizer is female  serves as a cue for various communities such as open source organizations, software development companies etc. to increase the proportion of females in leadership roles in order to encourage more female participation in their organizations. Practitioners can also aim to build tools such as~\cite{uddin2017automatic,uddin2017opiner} which can perform automated analysis and summarization of discussion and events in relevant Meetup groups. Such tools can help software developers to gain knowledge and discover information which might not be available on other channels.

\vspace{-.1cm}
\section{Conclusion and Future Work}
\label{sec.conclusion}
In this paper, we performed an empirical analysis of the events  organized by Meetup groups related so software development. To the best of our knowledge we are the first to look at the Meetup groups and events in detail, from a software engineering perspective. We first randomly sampled 452 events from a candidate set of 185,758 events organized by groups related to software development. We then did a qualitative analysis of these 452 events using open coding procedure and subsequent labeling. After we developed the categories we analyzed the popularity of the event categories, based on how often they are organized. We found that categories {\em Talks by Domain Experts, Hands-on Sessions, and Open Discussions} are the most popular. We also did a gender based diversity analysis on members of Meetup groups and found that 19.82\% of members are female on average, which is a higher proportion as compared to female participation in other social channels related to software development~\cite{vasilescu2015gender,david2008community,vasilescu2013gender}. Also based on Meetup topics we were able to build an association graph of Meetup topics, which we hope will help the researchers as well as the software community to understand and appreciate the various inter relationships in social interactions for software development.



Our future work is an expansive goal to combine data from online social groups such as  \textit{Facebook}, \textit{Reddit}, \textit{Twitter}, \textit{Hackernews} etc. with the data from offline social groups from resources such as \textit{Meetup}, \textit{EventBrite} etc. for further detailed analysis. The combined data can be used to compare and contrast the strengths and weaknesses of such online and offline interactions and leverage these to aid and assists in software development tasks. We also  plan to analyze in detail what makes an event or a group popular among software developers and if the topics of discussion differ in on-line and off-line communities. This can be based on the data based approach followed in~\cite{lai2014can,pramanik2016can}, which can be complemented by actually doing a survey with the members querying them why they prefer attending some events while skipping others. Another analysis that we aim to perform is to understand the reasons on the reasons that affect female participation in Meetup groups. The insights gathered from such works can help organizers  better manage their Meetup groups and events, as well as have an increased participation from various diverse populations. The identification of categories of events done in this work  as well as the analysis of topic popularity is a first step towards the accomplishment of the bigger goal of understanding the mechanisms by which software development communities thrive in offline and online settings and share information with one another.

\bibliographystyle{ACM-Reference-Format}
\bibliography{icse19_meetup_short}

\end{document}